\documentclass[twoside,12pt,CJK]{article}  %%%%% This command is necessary for LaTeX file
\usepackage{indentfirst} %%%% This package causes indentation of the first line after \section command
\usepackage{bm} %%%%% This package id used for typesetting bold italic math. letters and symbols
\usepackage{graphicx} %%% This package is used for inserting eps figures in body text (method 1)
\usepackage{amsmath} %%%%% This package is used for typesetting complicated math. formulae
\usepackage{epsfig}
\begin{document} %%%% Begin the LaTeX document file. This command is necessary for LaTeX files

%\begin{CJK*}{GBK}{kai}  %%%% begin Chinese, Japanese, and Korea languages environment

%-------------------  First Head  -----------------------------------------

%=================== Text begin here =============================================

\begin{center}
\LARGE\bf A single diffractive optical element for implementing spectrum-splitting and beam-concentration functions simultaneously with high diffraction efficiency$^{*}$    %% article title
\end{center}

\begin{center}  %% authors
Jia-Sheng Ye$^{\rm a,b}$, \ \ Jin-Ze Wang$^{\rm c}$,
\ Qing-Li Huang$^{\rm c}$, \ Bi-Zhen Dong$^{\rm c}$,
\ Yan Zhang$^{\rm a,b}$, \ and  \ Guo-Zhen Yang$^{\rm
c}$
\end{center}

\begin{center}  %% address or affiliation
\begin{small} \sl
${}^{\rm a)}$ Department of Physics, Capital Normal University,
Beijing 100048, P. R. China\\   %%%% address a)
${}^{\rm b)}$ Beijing Key Lab for THz Spectroscopy and Imaging, Key
Lab of THz Optoelectronics, Ministry of Education, Beijing 100048,
P. R. China\\   %%%% address b)
${}^{\rm c)}$ Institute of Physics, Chinese Academy of Sciences,
Beijing 100190, P. R. China\\   %%%% address c)
\end{small}
\end{center}

\vspace*{2mm}

\begin{center}  %% abstract
\begin{minipage}{15.5cm}
\parindent 20pt\small
In this paper, a novel method is proposed, and employed to design a
single diffractive optical element (DOE) for implementing
spectrum-splitting and beam-concentration (SSBC) functions
simultaneously. We develop an optimization algorithm, through which
the SSBC DOE can be optimized within an arbitrary thickness range,
according to the limitations of modern photolithography technology.
Theoretical simulation results reveal that the designed SSBC DOE has
a high optical focusing efficiency. It is expected that the designed
SSBC DOE should have practical applications in high-efficiency solar
cell systems.
\end{minipage}
\end{center}

\begin{center}  %% keywords
\begin{minipage}{15.5cm}
\begin{minipage}[t]{2.3cm}{\bf Keywords: }\end{minipage}
\begin{minipage}[t]{13.1cm}
diffractive optical element, spectrum-splitting and beam-concentration functions,
thickness optimization algorithm, high optical focusing efficiency,
solar cell systems%%%%% key words go here

\end{minipage}\par\vglue8pt
{\bf PACS:}
42.15.Eq, % Optical system design
42.25Fx, %Diffraction and scattering
%42.25Hz, %Interference
42.79Ek %Solar collectors and concentrators

%%% PACC codes go here

\end{minipage}
\end{center}

%%%%%%%% footnote for foundation information and E-mail address for corresponding author
\footnotetext[1]{Project supported by National Basic Research
Program of China (Grant 2011CB301801) and the National Natural
Science Foundation of China (Grants 91233202, 10904099, 11204188,
61205097 and 11174211).} \footnotetext[2]{Corresponding authors.
E-mail: yzhang@mail.cnu.edu.cn(Yan Zhang);
yanggz@aphy.iphy.ac.cn(Guo-Zhen Yang)}

\section{Introduction}  %%%  section command 1.
Energy and environmental problems are common challenges we are all
facing in the world. Solar energy, due to its clearness and huge
amount, fuels the need for the future. Nowadays, the solar energy is
usually utilized through the solar cells due to the photovoltaic
(PV) effect \cite{Razykov:SE2011}. The solar cell is composed of two
parts, i.e., the optical system and the cell system. The ``merit" of
a PV  cell depends upon the manufacturing cost of the PV material as
well as the cell's efficiency.

In order to reduce the cost, the incident sunlight is often highly
concentrated to offset the expensive cell \cite{Chang:USpatent2011}.
An alternative way is decreasing the fabrication difficulty by using
an optical system. In this case, the sunlight is spatially separated
into different regions, where properly selected single junction
solar cells are aligned \cite{Imenes:2004}. Since cell stacking and
epitaxial technologies \cite{Kurtz:OE2010} are no longer applied, it
becomes much cheaper and easier.

On the other hand, for achieving a high cell efficiency, the
following three main strategies are used \cite{Barnett:PPRA2009}.
Firstly, high efficiency converting materials such as the compounded
III-IV materials are selected \cite{Tanabe:Energies2009}, instead of
silicon. Secondly, several junctions with different bandgaps are
united together from top to bottom \cite{Tobias:PP2002}, so as to
break the maximum conversion efficiency of a single bandgap
converter limited by the Shockley Read Hall equation
\cite{Shockely:JAP1961}. However, there exist some constraints on
choosing the materials and substrates due to the lattice matching
problems.  Thirdly, the cell efficiency may be also raised by using
some new mechanism, new materials, or subwavelength structures,
including local light trapping techniques
\cite{Rim:APL2007,Ferry:OE2010,Raman:OE2011,Zheng:CPB2011},
plasmonics \cite{Catchpole:OE2008,Atwater:NM2010,Ren:OE2011},
organic solar cells \cite{Hoppe:JMR2004}, photonic crystals
\cite{Chutinan:OE2009,Park:OE2009,Zheng:CPL2011}, optical waveguides
\cite{Ruhle:OE2008}, nano-patterned structures
\cite{Liu:OE2011,Jonsson:IJP2011,Xiong:CPB2008}.

On summarizing the above two aspects of a high conversion efficiency
and low cost, we can find two critical optics issues in promoting
wide applications of solar cells. One is the spectrum-splitting
problem, and the other is the beam-concentration problem. For
implementing the spectrum-splitting function, three types of
elements are usually applied, namely prisms, diffraction gratings,
and dichroics. For realizing the beam-concentration function, an
optical lens needs to be employed. In previous papers, these two
functions are generally realized by two elements in successive steps
\cite{Science:2007,Mccambridge:2011}. As a result, it brings about
such problems as system stableness, system compactness, and
alignment errors. To solve this problem, in this paper we propose to
design a single diffractive optical element (DOE) that can implement
spectrum-splitting and beam-concentration (SSBC) functions
simultaneously. In order to make use of the modern photolithography
technology in the fabrication process, an optimization algorithm is
developed so that the thickness of the DOE can be confined within an
arbitrary range.

This paper is organized as follows. In Section 2, the designing
principle of the SSBC DOE and the thickness optimization algorithm
are described in detail with formulas. In Section 3, the parameters
of the designed SSBC DOEs are given, and then their performance
results are presented with explanations. A conclusion is drawn in
Section 4 with some discussions.
\section{Designing process of the single diffractive optical element (DOE) that implements spectrum-splitting and beam-concentration (SSBC) functions simultaneously}  %%%  2.

\subsection{Designing principle of the SSBC
DOE}\label{subsec:bulkDOEdesign}  %%%  2.1.  subaection
\begin{figure}[htb]
\centerline{\includegraphics[width=10cm]{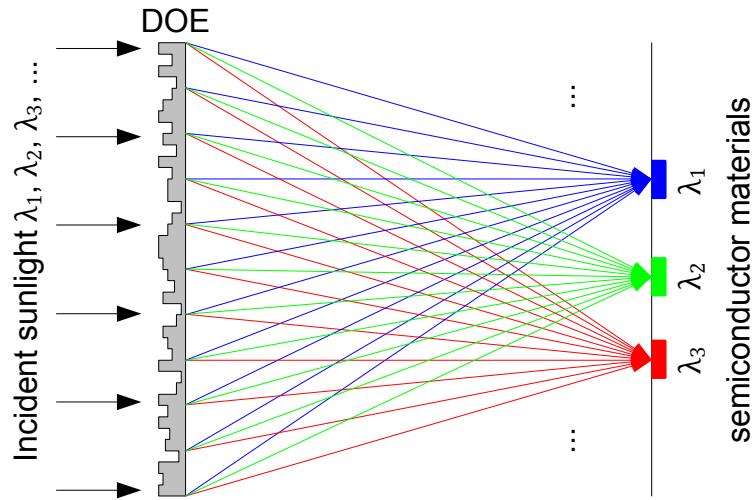}}
\caption{Schematic diagram of a single diffractive optical element
(DOE) that implements spectrum-splitting and beam-concentration
(SSBC) functions simultaneously.}\label{fig:schematic}
\end{figure}
Figure \ref{fig:schematic} depicts the schematic diagram of the
proposed single SSBC DOE. The sunlight is normally incident upon the
front surface of the DOE on the input plane. Due to the phase
modulation of its surface-relief profile, the sunlight with
different wavelengths will be separately focused at their
preassigned positions on the output plane, as shown in Fig.
\ref{fig:schematic}. If we put proper semiconductor materials at the
corresponding focal positions, the solar cell conversion efficiency
is expected to be significantly raised. Then, how to design such a
single SSBC DOE?

\begin{figure}[htb]
\centerline{\includegraphics[width=10cm]{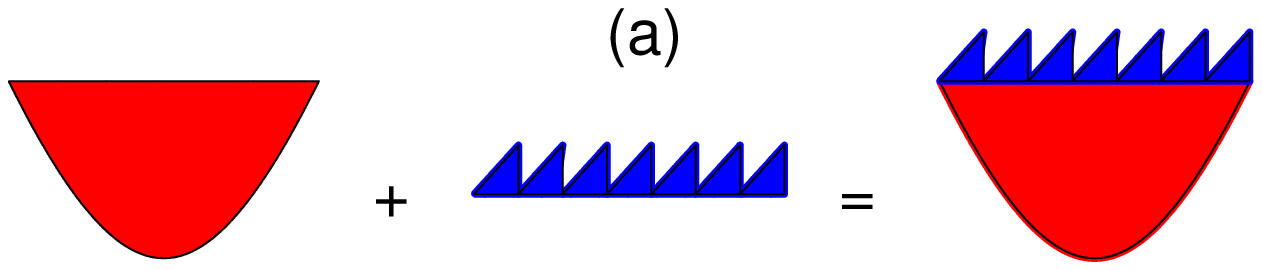}}\vskip0.3cm
\centerline{\includegraphics[width=10cm]{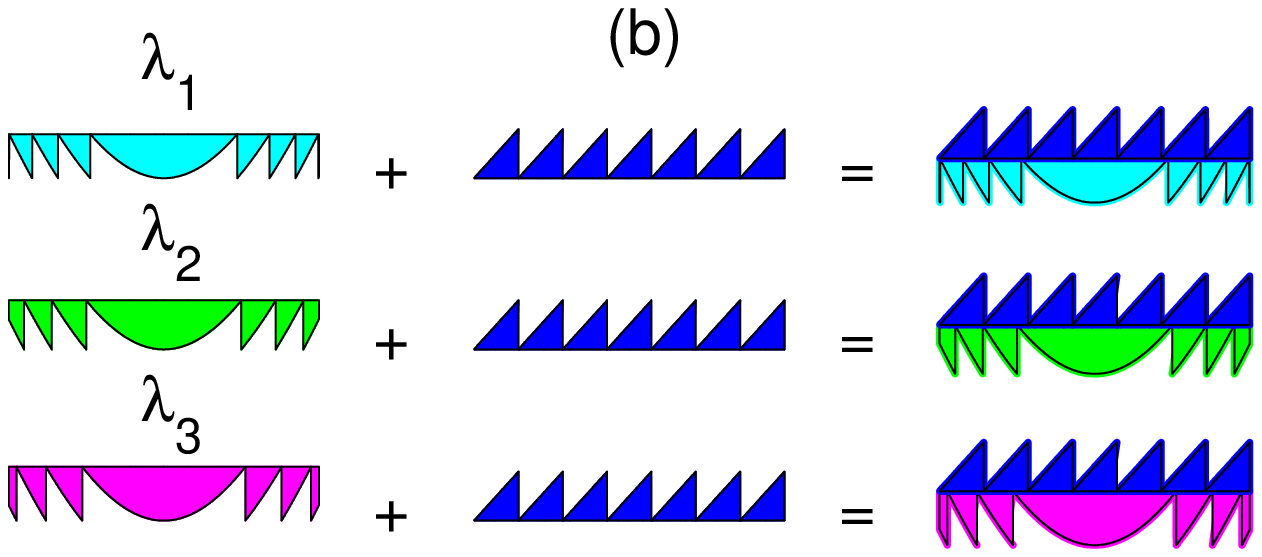}}
\caption{Diagram for interpreting the designing principle of the
SSBC DOE.}\label{fig:DOE}
\end{figure}
As is well known, the diffraction grating and lens can realize
spectrum-splitting and beam-concentration functions, respectively.
From our experience in physical optics, if we attach these two
elements together, the combined element should have both functions
simultaneously, as shown in Fig. \ref{fig:DOE}(a). From diffractive
optics, for a definite incident wavelength $\lambda_i$, the focusing
refractive lens in Fig. \ref{fig:DOE}(a) has a thickness function of
\begin{equation}
h_{\mathrm{rl}}=-x^2/\{2\times[n(\lambda_i)-1)]\times
f\}\,,\label{eq:reflens}
\end{equation}
where $n(\lambda_i)$ represents the refractive index of the lens at
wavelength $\lambda_i (i=1,2,3,...)$; $x$ stands for the horizontal coordinate
with origin at the center; $f$ denotes the preset focal length. For
the spectrum-splitting grating, it may be regarded as a Fresnel
prism, as shown by the blue element in Fig. \ref{fig:DOE}(a). Its
thickness is given by
\begin{equation}
h_{\mathrm{g}}=\mathrm{Mod}(px,\Delta h_0)\,,\label{eq:grating}
\end{equation}
where $\mathrm{Mod}(A,B)$ means the modulus function and its value
situates within $[0,B)$ for any positive real numbers $A$ and $B$;
$p$ denotes the slope of the prism; $\Delta
h_0=\lambda_0/[n(\lambda_0)-1]$, and $\lambda_0$ is chosen as the
average wavelength in the considered wavelength range.
Theoretically, although the combined element can fulfill the SSBC
functions, it usually has a too large thickness to fabricate because
of the thick refractive lens, as shown by the red element in Fig.
\ref{fig:DOE}(a). Therefore, we call it as a bulk SSBC DOE, whose
thickness is given by
\begin{equation}
h_\mathrm{b}=h_{\mathrm{rl}}+h_\mathrm{g}\,.\label{eq:DOE}
\end{equation}

In order to utilize the modern photolithography technology, the
thickness of the designed SSBC DOE should be confined in a
reasonable range. In physical optics, as the 2$\pi$ phase can be
added or removed without influencing the performance of the DOE, the
refractive lens may be attenuated into a Fresnel lens while the
grating part is kept unchanged. For instance, at wavelength
$\lambda_1$, the thickness of the combined element after attenuation
is written as
\begin{equation}
h_{11}(\lambda_1)=\mathrm{Mod}(h_{\mathrm{rl}},\Delta
h_1)+h_{\mathrm{g}}\,, \label{eq:thickness}
\end{equation}
where $\Delta h_1=\lambda_1/[n(\lambda_1)-1]$, corresponding to a
phase change of $2\pi$ for wavelength $\lambda_1$; $n(\lambda_1)$
denotes the material refractive index at wavelength $\lambda_1$.
When the incident wavelength becomes $\lambda_2$, $\lambda_3$,...,
we can obtain $h_{12}$, $h_{13}$,..., as shown in Fig.
\ref{fig:DOE}(b). So far, a new problem rises. For the solar cell
system, as the incident sunlight includes multiple wavelengths, from
Eq. (\ref{eq:thickness}), we can not get a unique boundary profile
of the designed SSBC DOE.
%\subsubsection{} %%% 2.1.1.  subsubsection
\subsection{Thickness optimization algorithm for designing the SSBC DOE}
\begin{figure}[htb]
\centerline{\includegraphics[width=13cm]{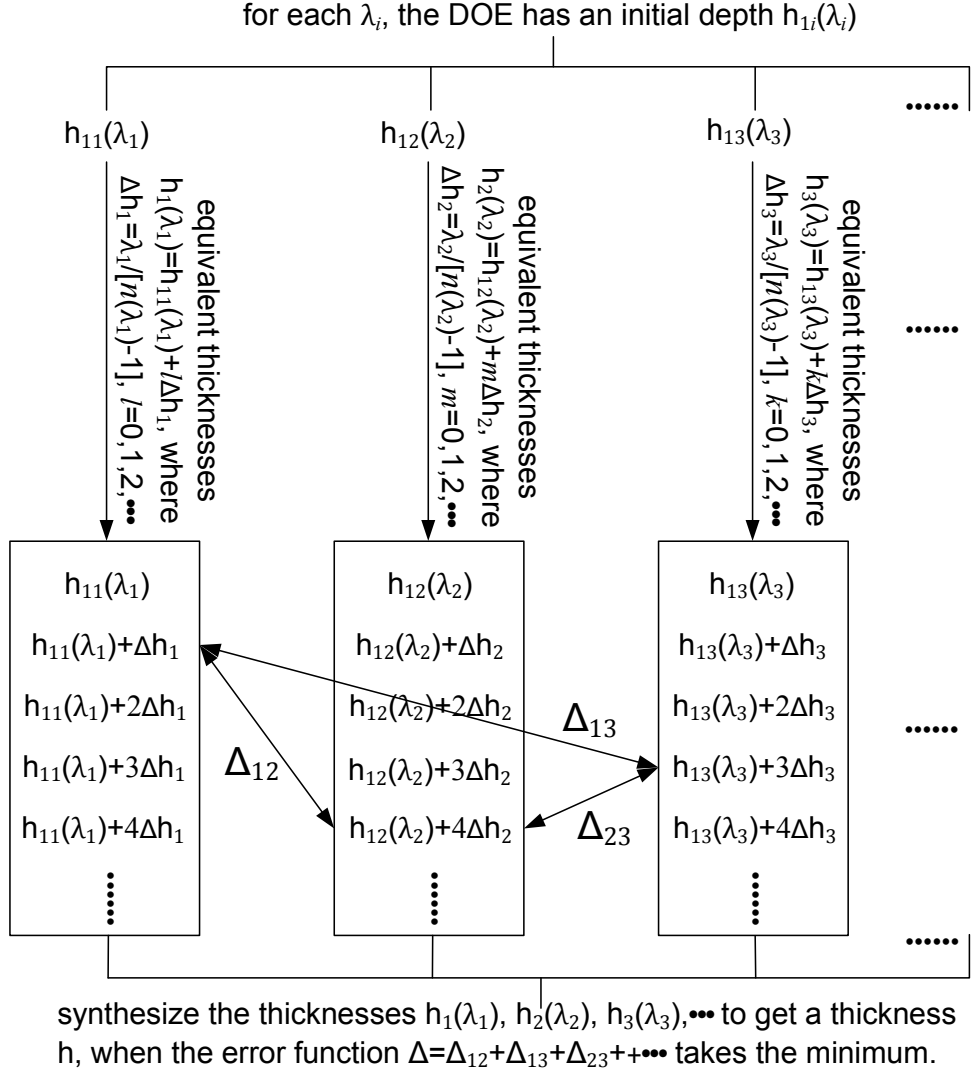}} \caption{Flow
chart for the thickness optimization
algorithm.}\label{fig:flowchart}
\end{figure}
To solve the above problem, in this subsection we develop a
thickness optimization algorithm. For wavelength $\lambda_1$, it has
a ground thickness $h_{11}$. Let's assume that the largest permitted
thickness of the designed DOE is $h_{\mathrm{max}}$, which is
usually limited by the experimental fabrication conditions. Since multiples of
2$\pi$ phase do not exert any impact on the DOE's
performance, a series of equivalent thicknesses may be obtained as
\begin{equation}
h_1=h_{11}+l\Delta h_1\,, \label{eq:wave1}
\end{equation}
where $l=0,1,2,...$, as shown in Fig. \ref{fig:flowchart}. The
maximum value of $l$ is governed by $h_1\leq h_{\mathrm{max}}$.
Similarly, for another incident wavelength $\lambda_2$, we obtain
its corresponding series of equivalent thicknesses as
\begin{equation}
h_2=h_{12}+m\Delta h_2\,, \label{eq:wave2}
\end{equation}
where $m=0,1,2,...$; $\Delta h_2=\lambda_2/[n(\lambda_2)-1]$, and
$n(\lambda_2)$ represents the material refractive index at
wavelength $\lambda_2$, as shown in Fig. \ref{fig:flowchart}.  The
maximum value of $m$ is determined by $h_2\leq h_{\mathrm{max}}$.

Following the above steps, for each wavelength we can generate its
unique set of thicknesses. From each set, we arbitrarily select a
thickness. If the total wavelength number is $M$, altogether we will have $M$
thicknesses. Thus the error between arbitrary two thicknesses is calculated
 as $\Delta_{ij}=|h_i-h_j|$. We define an error function as
\begin{equation}
\Delta=\sum_{i=1,M}^{j=i+1,M}\Delta_{ij}\,. \label{eq:error}
\end{equation}
Through searching the minimum value of $\Delta$, the optimum
thickness in each set will be determined, as seen in Fig.
\ref{fig:flowchart}. Finally, the thickness of the SSBC DOE is
optimized as
\begin{equation}
h=\frac{1}{M}\sum_{i=1}^Mh_i\,.\label{eq:DOEthickness}
\end{equation}

As the thickness $h$ in Eq. (\ref{eq:DOEthickness}) may take any
continuous value, in the following we refer to it as
$h_\mathrm{c}(=h)$. Since modern photolithographic fabrications are
implemented through binary masks, the continuous DOEs are usually
quantized into multilevel ones in the following way. If the maximum
permitted thickness is $h_{\mathrm{max}}$ and the total quantization
level number is $N$, each step depth is $\Delta
h=h_{\mathrm{max}}/N$. The thickness of the quantized DOE
($h_\mathrm{q}$) is derived from that of the continuous one
$(h_\mathrm{c})$ as
\begin{equation}
h_\mathrm{q}=\mathrm{Int}(h_\mathrm{c}/\Delta h)\times\Delta h\,,
\end{equation}
where $\mathrm{Int}(C)$ means taking the maximum integer no larger than a real number $C$.

\section{Performance simulation results of the designed SSBC DOEs}
\subsection{Designing parameters and the performance characterization}
To demonstrate the validity of the proposed method, we will design
the SSBC DOE and characterize its performance. Parameters are
selected as follows. The incident wavelength ranges from 0.45 to
0.65$\mu\mathrm{m}$. For the designing processes in Figs.
\ref{fig:DOE} and \ref{fig:flowchart}, we choose three wavelengths
as 0.45, 0.55, and 0.65$\mu\mathrm{m}$. Therefore, the wavelength
number in Eq. (\ref{eq:DOEthickness}) is $M=3$. The average
wavelength is $\lambda_0=0.55\mu\mathrm{m}$. The distance between
the input plane and the output plane is $f=800\mathrm{mm}$. We
define a length unit as $l_0=\sqrt{\lambda_0\times
f}\approx0.6633\mathrm{mm}$. The grating angle is $\theta=\pi/160$,
therefore, the constant $p$ in Eq. (\ref{eq:grating}) equals
$p=\tan(\pi/160)$. Both the input and output planes have the same
sizes of $32l_0\approx21.23\mathrm{mm}$. The input and output planes
are equally quantized into 4096 segments, with each element scaling
about 5.18$\mu\mathrm{m}$. The material of the DOE is chosen as
fused silica, whose refractive index changes slightly in the visible
region. Hence, we neglect the dispersive effect and assume the
refractive index to be a constant as $n=1.46$ \cite{Bass:1995}.
Following the above-mentioned designing procedure in Section 2, the
thickness of the SSBC DOE can be obtained.

After designing the DOE, under thin-element approximation we can
calculate the field distribution on the input plane (just passing
through the DOE) as
\begin{equation}
U_1(X_1,\lambda_i)=\exp[j\frac{2\pi(n-1)\times
h(X_1)}{\lambda_i}]\,,
\end{equation}
where $h(X_1)$ represents the thickness of the DOE; $h(X_1)$ equals
$h_\mathrm{c}(X_1)$ or $h_\mathrm{q}(X_1)$ for the bulk or the
optimized SSBC DOE, respectively; $X_1$ denotes the transverse
coordinate of the sampling point on the input plane. It is noted
that the reflection losses on both interfaces of the DOE are
ignored. As the propagation distance is far enough, the Fresnel
diffraction integral is applied to calculating the field on the
output plane as \cite{Goodman:1968}
\begin{equation}
U_2(X_2,\lambda_i)=\int
U_1(X_1,\lambda_i)\hat{G}(X_1,X_2,\lambda_i)\mathrm{d}X_1\,,
\end{equation}
where $X_2$ represents the transverse coordinate on the output plane;
$\hat{G}$ denotes the Fresnel diffraction integral kernel, as
given by
\begin{equation}
\hat{G}(X_1,X_2,\lambda_i)=\frac{1}{\sqrt{j\lambda_i f}}\exp(j2\pi
f/\lambda_i)\exp[j\pi (X_1-X_2)^2/(\lambda_if)].
\end{equation}
In order to characterize the performance of the DOE, the optical
focusing diffraction efficiency $\eta_i$ for each wavelength
$\lambda_i$ is defined as
\begin{equation}
\eta_i=\frac{\int_{x_i-l_0}^{x_i+l_0}|U_2(X_2,\lambda_i)|^2\mathrm{d}X_2}{\int_{-\infty}^{+\infty}
|U_1(X_1,\lambda_i)|^2\mathrm{d}X_1}\,, \label{eq:efficiency}
\end{equation}
where $x_i$ denotes the geometrical transverse focal position for
wavelength $\lambda_i$; the symbol $|U_j| (j=1,2)$ represents the
magnitude of a complex number $U_j$. The ultimate optical focusing
diffraction efficiency of the designed SSBC DOE is averaged as
\begin{equation}
\eta=\frac{1}{M}\sum_{i=1}^{M}\eta_i\,.
\end{equation}

\subsection{Performance results of the designed bulk SSBC DOE}\label{sec:bulkDOE}
\begin{figure}[htb]
\centerline{\includegraphics[width=9cm]{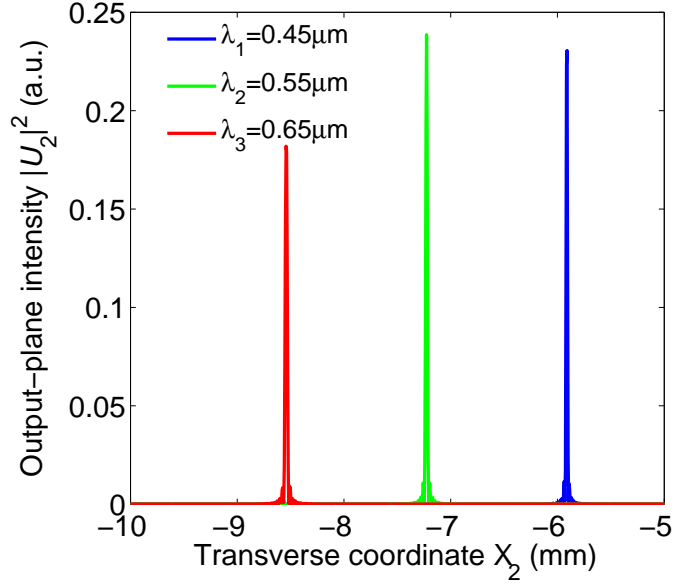}}
\caption{Normalized intensity distributions of the designed bulk
SSBC DOE on the output plane. }\label{fig:bulklens}
\end{figure}
Firstly, we design the bulk SSBC DOE, which consists of the
refractive lens and the diffraction grating without thickness
limitation. By following the designing procedure in subsection 2.1,
the thickness of the bulk SSBC DOE is evaluated. Apparently, on
ignoring the material dispersive effect, the boundary of the bulk
SSBC DOE does not depend on the designing wavelength. Therefore, it
suits all the designed wavelengths well and is expected to have very
good performance. Through calculating the Fresnel diffraction
integral, the intensity distributions on the output plane are
obtained and displayed in Fig. \ref{fig:bulklens}. The blue, green,
and red curves correspond to the intensity distributions for
incident wavelengths of 0.45, 0.55, and 0.65$\mu\mathrm{m}$,
respectively. It is seen from Fig. \ref{fig:bulklens} that the light
is separately concentrated very well on the output plane. Numerical
results reveal that the focusing diffraction efficiency reaches
 $87.93\%$. The full-width-at-half-maximum (FWHM) dimensions are
15.55, 20.73, and 25.91$\mu\mathrm{m}$ for wavelengths of 0.45,
0.55, and 0.65$\mu\mathrm{m}$, respectively.  The corresponding
transverse real focal positions are -5.91, -7.23, and
-8.54$\mathrm{mm}$. As a comparison, the preset focal positions are
also evaluated from the geometrical formula
$x_i=-f\times\tan\{\arcsin[(n-1)\times\sin(\theta)]\times\lambda_i/\lambda_0\}$
as -5.91, -7.23, and -8.54$\mathrm{mm}$. It is concluded that the
designed bulk SSBC DOE has realized both spectrum-splitting and
beam-concentration functions very well at the preassigned positions.

\subsection{Performance results of the optimized SSBC DOE}\label{sec:diffDOE}
\begin{figure}[htb]
\centerline{\includegraphics[width=9cm]{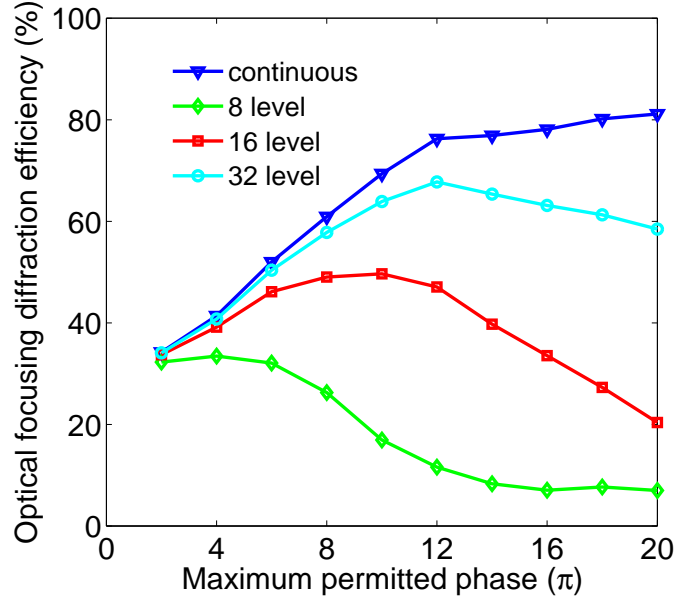}}
\caption{Optical focusing diffraction efficiency versus the maximum
permitted phase for the optimized SSBC DOEs.}\label{fig:efficiency}
\end{figure}
In subsection 3.2, although we have successfully designed the bulk
SSBC DOE with high diffraction efficiency, the total thickness of
the designed DOE is as large as $153.82\mu\mathrm{m}$. In fact, such
a thick element is very difficult for photolithographic fabrication
of fused silica. In this subsection, the thickness optimization
algorithm is adopted to attenuate the thickness. We have calculated
the diffraction efficiency of the optimized SSBC DOEs on varying the
maximum permitted phase, as shown in Fig. \ref{fig:efficiency}. The
phase is involved with the average wavelength, which means that
2$\pi$-phase increase leads to a thickness increment of $\Delta
h_0=\lambda_0/(n-1)\approx1.20\mu\mathrm{m}$. The blue curve
corresponds to the optimized SSBC DOEs with a continuous phase. From
Fig. \ref{fig:efficiency}, it is obvious that the diffraction
efficiency of the continuous DOE is monotonically increased as the
maximum permitted phase is enlarged. It can be understood. With the
increase of the maximum permitted phase, because we have more
choices in each set of thicknesses, it will be easier to find a
thickness that satisfies all the designed wavelengths better.

Considering practical fabrication of the optimized SSBC DOE, the
continuous DOE is quantized to have a multilevel profile. The green,
red, and cyan curves correspond to the 8-level, 16-level, and
32-level DOEs, respectively. For the designed multilevel DOEs, when
the quantization level number is increased, the focusing diffraction
efficiency is increased. It is due to the decrease of the
quantization error. For a given quantization level number, there
exists an optimum maximum permitted phase. It can be explained as
follows. When the maximum permitted phase is too small, the
selection in each thickness set is strictly limited and thus the
optimized thickness $h_\mathrm{q}$ deviates very far from the
selected thickness $h_i$ for each wavelength. On the other hand,
when the maximum permitted phase is too large, it will be much
easier to find an accurate thickness $h_\mathrm{c}$. However, for a
given quantization level number each step depth is very large, which
results in a large boundary quantization error.

\vskip0.3cm
\begin{figure}[htb]
\centerline{\includegraphics[width=8.5cm]{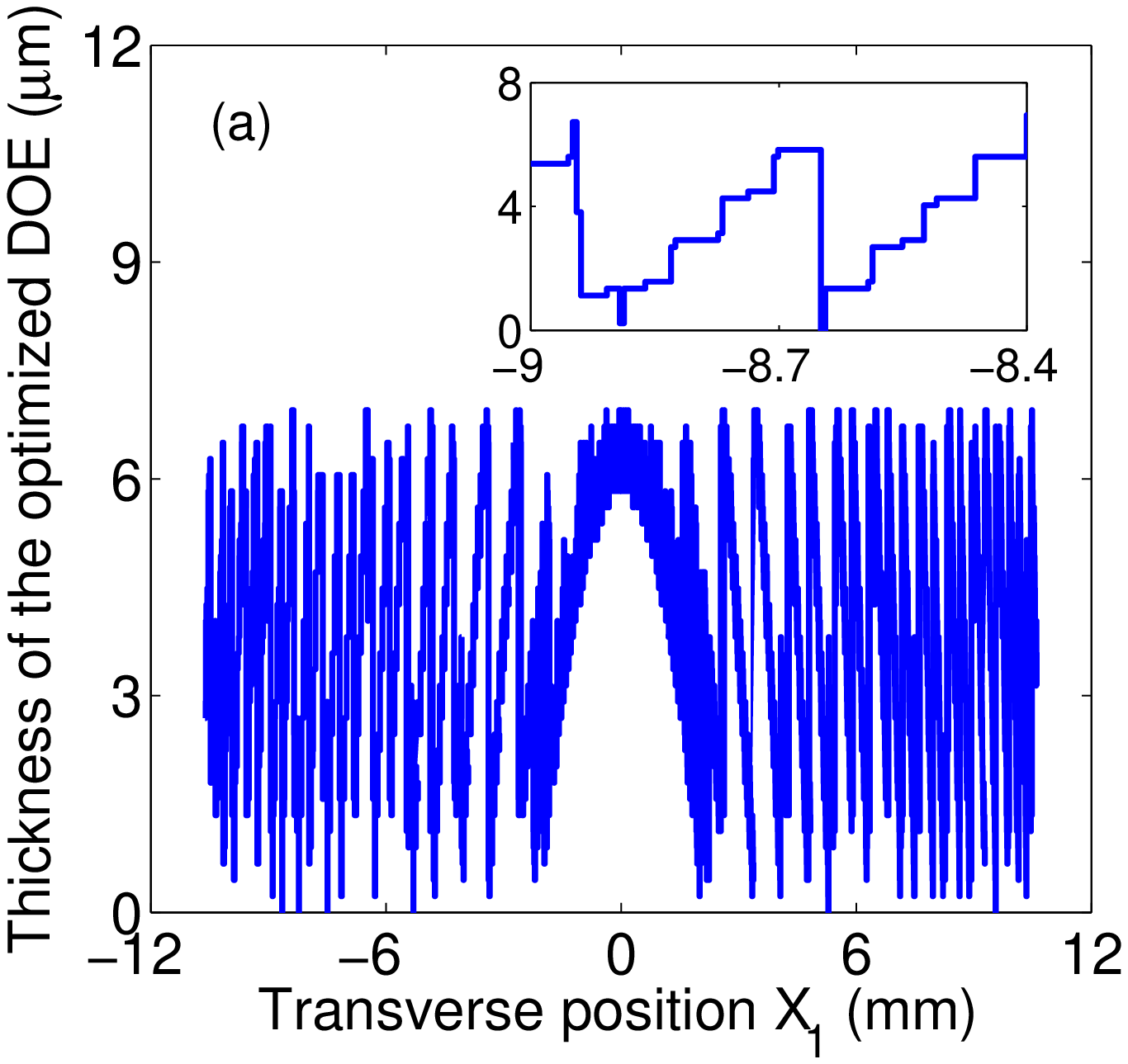}}\vskip0.5cm
\centerline{\includegraphics[width=8.5cm]{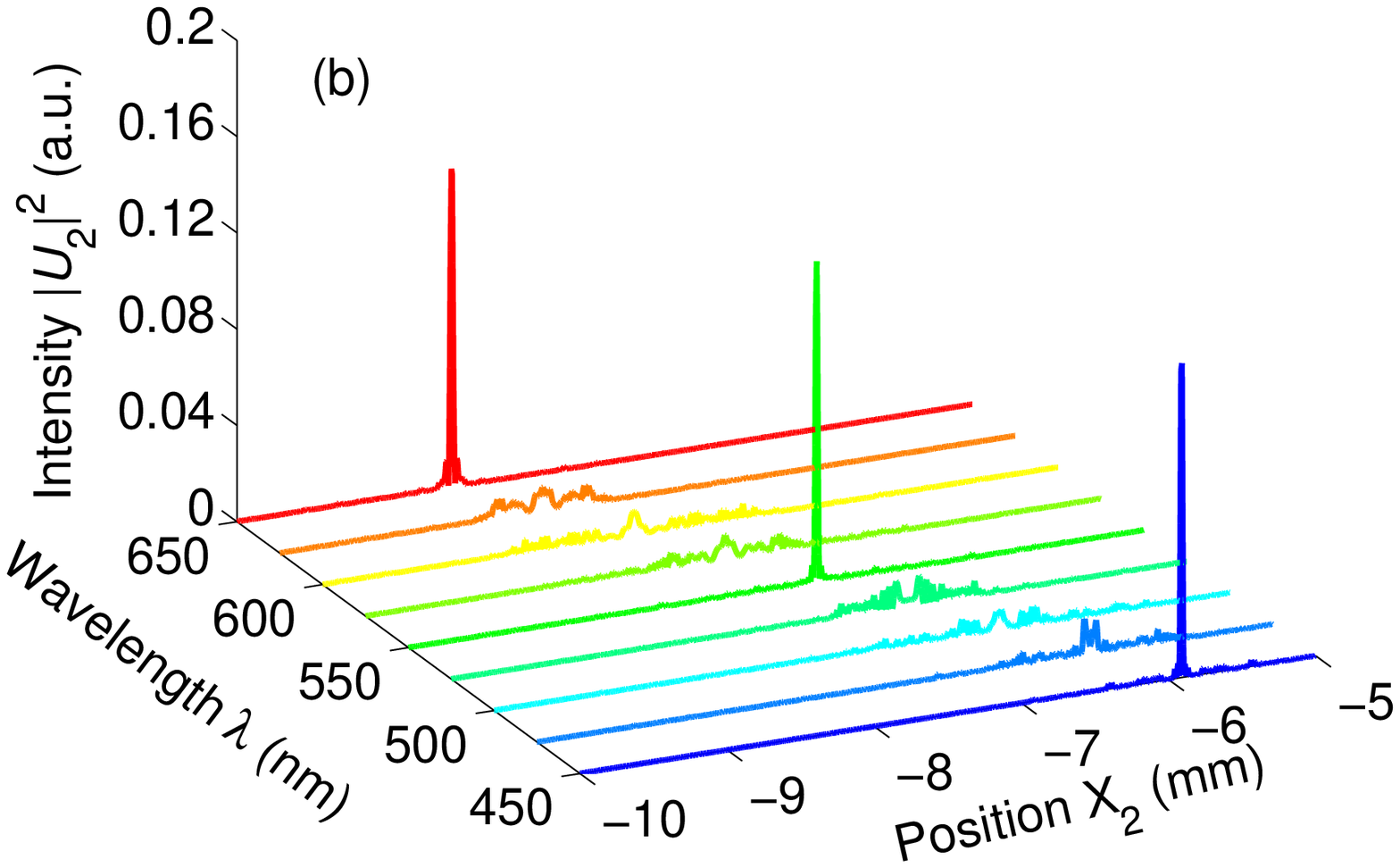}} \caption{(a)
Boundary profile of the optimized 32-level SSBC DOE. (b) Normalized
output-plane intensity distribution of the optimized 32-level SSBC
DOE with the maximum permitted phase of 12$\pi$.
}\label{fig:difflens}
\end{figure}

Figure \ref{fig:difflens}(a) shows the boundary profile of the
optimized 32-level SSBC DOE, which basically looks like a Fresnel
lens. The inset figure magnifies the local part of the DOE boundary.
For the optimized 32-level SSBC DOE, numerical results reveal that
the optimum maximum permitted phase is 12$\pi$, corresponding to the
maximum thickness of
$12\pi\times[\lambda_0/(n-1)]/(2\pi)\approx7.17\mu\mathrm{m}$. Under
illuminating wavelengths of 0.45, 0.55, and 0.65$\mu\mathrm{m}$, the
intensity distributions on the output plane is plotted in Fig.
\ref{fig:difflens}(b) by the blue, green, and red curves,
respectively. It is clearly demonstrated that the optimized 32-level
SSBC DOE has good focusing properties for all the three designed
wavelengths, as most of the incident power is concentrated in the
mainlobe and the side lobes are suppressed well. Numerical results
reveal that the average optical focusing diffraction efficiency is
$67.75\%$ for these three wavelengths. The three FWHM sizes are
15.55, 20.73, and 25.91$\mu\mathrm{m}$ for incident wavelengths of
0.45, 0.55, and 0.65$\mu\mathrm{m}$, respectively. In addition, the
transverse preset and real focal positions are the same to be -5.91,
-7.23, and -8.54$\mathrm{mm}$.

Since the sunlight has a continuous spectrum, we are curious about
how the optimized 32-level SSBC DOE in Fig. \ref{fig:difflens}(a)
performs for other wavelengths? For this reason, we choose six
incident wavelengths (475, 500, 525, 575, 600 and 625nm) to
calculate the output intensity, as displayed in Fig.
\ref{fig:difflens}(b). For these six wavelengths, although the
intensity peaks are not high, the energy concentration regions
situate in turn with the increase of the incident wavelength. Even
if more wavelengths are considered, we can draw a similar
conclusion. In solar cell applications, we are more interested in
the energy concentration position and efficiency, rather than a
higher intensity peak. Numerical results reveal that the
corresponding six diffraction efficiencies are 54.61\%, 49.03\%,
69.16\%, 60.03\%, 56.08\% and 70.40\%. Above all, it is concluded
that the optimized 32-level DOE not only realizes the expected SSBC
functions, but also has a high focusing diffraction efficiency.

\section{Conclusions and discussions}
In this paper, we propose a new method for designing the single DOE
with SSBC functions. The main novelties include these four points.
Firstly, we present an analytical thickness formula for the bulk
SSBC DOE. Secondly, a thickness optimization algorithm is developed,
so that the SSBC DOE can be optimized within an arbitrary thickness
range. Thirdly, theoretical simulations demonstrate that the
optimized SSBC DOE has a high optical focusing efficiency in the
light concentration region. The last but not least, when the
thickness of the SSBC DOE is limited in a reasonable scope, it may
be fabricated by the modern photolithography technology. More
importantly, if a hard mold with a complementary structure is
manufactured, mass amount of the SSBC DOEs can be duplicated by
using the micro imprint technology. Therefore, it probably provides
a cheap way for high-efficiency solar cells.

In this paper, we just select several wavelengths in the visible
region instead of the total solar band, because of the following two
reasons. One reason is that our main purpose is to demonstrate the
validity of the proposed method, and simultaneously the visible
range is an important band in the solar spectrum. The other reason
is that experimental verifications will be easier to be carried out
for the visible light. Fabrications of the optimized SSBC DOE are in
progress and their performance will be measured to verify our
theoretical predictions.

%%%% References typesetting

%\end{CJK*}  %% end Chinese, Japanese, and Korea languages environment
\end{document}